# Spin Seebeck Effect in Asymmetric Four-Terminal Systems with Rashba Spin-Orbit Coupling

Jun Zhou,[1*] Biao Wang,[1] Mengjie Li,[1]

Tsuneyoshi Nakayama,[1,2] and Baowen Li[1,3,4*]

[1]Center for Phononics and Thermal Energy Science, School of Physics Science and Engineering, Tongji University, Shanghai 200092, People's Republic of China

[2]Hokkaido University, Sapporo 060-0826, Japan

[3]Department of Physics, Center for Computational Science and Engineering, and Graphene Research Center, National University of Singapore, Singapore 117546, Republic of Singapore

[4]NUS Graduate School for Integrative Sciences and Engineering, National University of Singapore, Singapore 117456, Republic of Singapore

**Abstract**

We propose a new type of the spin Seebeck effect (SSE) emerging from the Rashba spin-orbit coupling in asymmetric four-terminal electron systems. This system generates spin currents or spin voltages along the *longitudinal* direction parallel to the temperature gradient in the absence of magnetic fields. The remarkable result arises from the breaking of reflection symmetry along the transverse direction. In the meantime, the SSE along the transverse direction, so-called the *spin Nernst* effect, with spin currents or spin voltages perpendicular to the temperature gradient can be simultaneously realized in our system. We further find that it is possible to use the temperature differences between four leads to tune the spin Seebeck coefficients.

* Authors to whom correspondence should be addressed. E-mail: zhoujunzhou@tongji.edu.cn, phononics@tongji.edu.cn

Spin caloritronics, the study of relationship between spin current and heat flow in magnetic materials, has attracted much attention in recent years [1-10], in particular after the discovery of the spin Seebeck effect (SSE) in metallic ferromagnets by Uchida *et al.* [1]. Analogous to the conventional charge Seebeck effect generating electric voltages, the SSE refers to the generation of spin currents (or spin voltages) in the presence of temperature gradient [11]. In general, the SSE can be divided into two categories [12]: i) the longitudinal SSE in which spin currents are generated parallel to the temperature gradient, ii) the transverse SSE, also called the spin Nernst effect, in which spin currents are generated perpendicular to the temperature gradient. The transverse SSE was first observed in metallic ferromagnetic films [1, 13], ferromagnetic semiconductors [4], and magnetic insulators [3] by using the inverse-spin-Hall effect. Later on, the longitudinal SSE has also been confirmed experimentally in magnetic insulators [14,15]. All of these experiments on the SSE have been performed for magnetic materials, in which the spin current is carried by magnon excitations.

In non-magnetic materials, the spin current is carried by conduction-electrons rather than magnons, for which there has been no experimental work on the SSE. Theoretically, Lü *et al.* [16] and Cheng *et al.* [17] have investigated the SSE in non-magnetic materials, two-dimensional (2D) mesoscopic electron systems, in the presence of both spin-orbit coupling (SOC) and magnetic fields. In the absence of magnetic fields, the transverse SSE analogous to the spin-Hall effect has been studied [18]. Both the intrinsic transverse SSE due to the SOC [19,20] and the extrinsic transverse SSE due to the spin-dependent skew scatterings [21] have been theoretically investigated. However, there exists no theoretical treatment on the occurrence of the *longitudinal* SSE in the absence of magnetic fields.

In this Letter, we propose a system realizing the longitudinal SSE in the absence of magnetic fields and suggest a practical scheme for implementing such a device composed of 2D mesoscopic asymmetric four-terminals with the Rashba SOC. Several developments in the study of the spin-Hall effect and Nernst effect [22-25], though for symmetric systems, make such systems a practical possibility. Figure 1 shows the schematic diagram of a square region of width $L$ in $x$-$y$ plane connected with four ideal leads also of width $L$. $T_\mu$ is the temperature at lead $\mu$, where $\mu$=1,2,3,4. The temperatures at leads 1 and 3 are $T_1 = T + \gamma \Delta T$ and $T_3 = T - (1-\gamma) \Delta T$ with $0 \leq \gamma \leq 1$, respectively. The temperatures at leads 2 and 4 are $T$. We break the reflection symmetry along the transverse direction ($y$-direction) by introducing an isosceles triangle region as shown in Fig. 1. Such asymmetric samples have been fabricated by Matthews *et al.* [25] in studying the thermal rectification effect.

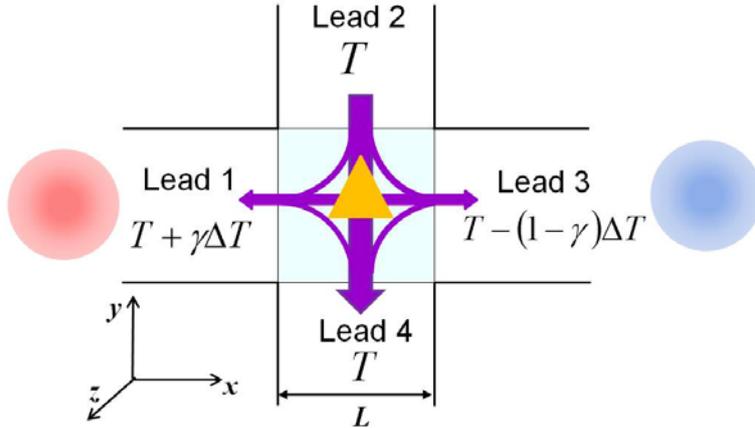

FIG. 1. (Color online) Schematic diagram of a 2D mesoscopic four-terminal electron system. The square region of width $L$ is connected to four leads with same width. The Rashba SOC works only in the square region. The isosceles triangle region breaks the reflection symmetry along $y$-direction. The temperature difference is applied between leads 1 and 3, where $T_1 = T + \gamma \Delta T$ and $T_3 = T - (1-\gamma)\Delta T$. The temperatures at leads 2 and 4 are taken as $T$. The spin currents (purple arrows) are generated in all leads.

The electric current $I_{\mu\sigma}$ into lead $\mu$ can be calculated from the Landauer-Büttiker formula [26,27]

$$I_{\mu\sigma} = \sum_{\nu\sigma'} L^{(0)}_{\mu\sigma,\nu\sigma'}(V_{\mu\sigma} - V_{\nu\sigma'}) + \sum_{\nu\sigma'} L^{(1)}_{\mu\sigma,\nu\sigma'}(T_\mu - T_\nu). \qquad (1)$$

Here we define the spin polarization $\sigma$ ($\sigma = \uparrow, \downarrow$) in $z$-direction. $L^{(n)}_{\mu\sigma,\nu\sigma'} = [e^{(2-n)}/h] \int dE \mathcal{T}_{\mu\sigma,\nu\sigma'}(E)(-\partial f_0/\partial E)[(E-E_F)/T]^n$ is the transport coefficient with $n = 0, 1$, in which $\mathcal{T}_{\mu\sigma,\nu\sigma'}(E)$ is the energy-dependent transmission coefficient from lead $\nu$ with spin $\sigma'$ (noted as $\{\nu\sigma'\}$) to lead $\mu$ with spin $\sigma$ (noted as $\{\mu\sigma\}$). $V_{\mu\sigma}$ ($V_{\nu\sigma'}$) is the spin-dependent potential related bias of $\{\mu\sigma\}$ ($\{\nu\sigma'\}$), and $f_0(E) = 1/\{\exp[(E-E_F)/(k_B T)] + 1\}$ [20,26]. We employ the following tight-binding Hamiltonian [28] to calculate the transmission coefficient

$$H = \sum_{i\sigma} \varepsilon_i c^\dagger_{i,\sigma} c_{i,\sigma} - t \sum_{\langle ij \rangle \sigma} c^\dagger_{i,\sigma} c_{j,\sigma}$$

$$+ t^{SO} \sum_i \left[ \left( c^\dagger_{i,\uparrow} c_{i+\delta_x,\downarrow} - c^\dagger_{i,\downarrow} c_{i+\delta_x,\uparrow} \right) - i \left( c^\dagger_{i,\uparrow} c_{i+\delta_y,\downarrow} + c^\dagger_{i,\downarrow} c_{i+\delta_y,\uparrow} \right) + \text{H. c.} \right], \qquad (2)$$

where $c^\dagger_{i,\sigma}$ and $c_{i,\sigma}$ are the creation and annihilation operators of electrons on site $i$ with spin $\sigma$. The onsite energy is $\varepsilon_i = 4t + V(i)$, where we choose $V(i) \to \infty$ when $i$ is inside the triangle region and $V(i) = 0$ in the rest region. The parameter $t = \hbar^2/2m^*a^2$ represents the hopping energy with the effective mass $m^*$ between the nearest-neighboring sites $\langle i,j \rangle$ of the lattice constant $a$. The parameter $t^{SO} = V_{SOC}/2a$ is expressed by the Rashba SOC constant $V_{SOC}$. $\delta_x$ and $\delta_y$ are the unit vectors along the *x*- and *y*-directions. The transmission coefficient $\mathcal{T}_{\mu\sigma,\nu\sigma'}(E) = \text{Tr}[\mathbf{\Gamma}_{\mu\sigma}\mathbf{G}^R\mathbf{\Gamma}_{\nu\sigma'}\mathbf{G}^A]$ can be calculated by the Green's function method [27,29], where $\mathbf{G}^R$ ($\mathbf{G}^A = [\mathbf{G}^R]^\dagger$) is the retarded (advanced) Green's function given by $\mathbf{G}^R = [E - \mathbf{H} - \sum_{\mu=1}^{4}(\mathbf{\Sigma}_\mu^R)]^{-1}$. Here $\mathbf{\Gamma}_{\mu\sigma} = i[\mathbf{\Sigma}_{\mu\sigma}^R - \mathbf{\Sigma}_{\mu\sigma}^A]$ with $\mathbf{\Sigma}_{\mu\sigma}^R$ being the self-energy, and $\mathbf{\Sigma}_{\mu\sigma}^A = [\mathbf{\Sigma}_{\mu\sigma}^R]^\dagger$. In the low-temperature limit $T \ll E_F/k_B$, the $E_F$-dependent transport coefficient can be expressed as [30]

$$L_{\mu\sigma,\nu\sigma'}^{(1)} = \frac{\pi^2 e k_B^2 T}{3h} \frac{d\mathcal{T}_{\mu\sigma,\nu\sigma'}(E)}{dE}\bigg|_{E=E_F}. \tag{3}$$

The spin Seebeck coefficient (SSC) for closed boundary (CB) conditions by setting the voltage as $V_{\mu\sigma} = 0$ for all $\{\mu\sigma\}$ and open boundary (OB) conditions by setting the spin current as $I_{\mu\sigma} = 0$ for all $\{\mu\sigma\}$ are written as

$$S_\mu^I = \frac{\hbar}{2e}\frac{I_{\mu\uparrow} - I_{\mu\downarrow}}{\Delta T}, \quad S_\mu^V = \frac{V_{\mu\uparrow} - V_{\mu\downarrow}}{2\Delta T}, \tag{4}$$

where $(I_{\mu\uparrow} - I_{\mu\downarrow})\hbar/2e$ is the spin current and $V_{\mu\uparrow} - V_{\mu\downarrow}$ represents the spin accumulation in lead $\mu$. $S_{\mu=1,3}^I$ ($S_{\mu=1,3}^V$) describe the longitudinal SSCs under CB (OB) conditions. Similarly, $S_{\mu=2,4}^I$ ($S_{\mu=2,4}^V$) describe the transverse SSCs under CB (OB) conditions (See Fig. 1). The substitution of the spin current $I_{\mu\sigma}$ of Eq. (1) into Eq. (4) yields the analytical expressions of $S_{\mu=1,3}^I$ with the definition of $\Delta_{\mu\nu} = \left[L_{\mu\uparrow,\nu\uparrow}^{(1)} + L_{\mu\uparrow,\nu\downarrow}^{(1)}\right] - \left[L_{\mu\downarrow,\nu\uparrow}^{(1)} + L_{\mu\downarrow,\nu\downarrow}^{(1)}\right]$ of the form

$$S_1^I = \frac{\hbar}{2e}[(\Delta_{12} + \Delta_{14})\gamma + \Delta_{13}], \quad S_3^I = -\frac{\hbar}{2e}[(\Delta_{32} + \Delta_{34})(1-\gamma) + \Delta_{31}]. \tag{5}$$

While, the analytical expressions of $S_{\mu=2,4}^I$ become

$$S_{\mu=2,4}^I = \frac{\hbar}{2e}[-\gamma\Delta_{\mu 1} + (1-\gamma)\Delta_{\mu 3}], \tag{6}$$

Equations (5) and (6) provide two features on the SSC: i) at low temperature $T \ll E_F/k_B$, both the longitudinal and the transverse SSCs show linear-temperature dependence. This is because $L^{(1)}_{\mu\sigma,\nu\sigma'}$ and $\Delta_{\mu\nu}$ are proportional to $T$ from Eq. (3), ii) the sharp energy-dependent transmission coefficient is required in order to obtain large SSC since $L^{(1)}_{\mu\sigma,\nu\sigma'}$ is proportional to the slope on $E$ of $\mathcal{T}_{\mu\sigma,\nu\sigma'}(E)$ as seen from Eq. (3).

In our numerical calculations, the width of square region ($L \times L$) is taken to be $L=20a=250$nm with $a=12.5$nm. The side length of the isosceles triangle region is 35nm and the bottom width is 50nm. We choose $t=5$meV and $T=0.58$K ($= 0.01t/k_B$) which ensures the electron transport is within the ballistic regime. The Rashba SOC constant $V_{SOC}$ is chosen to be $2.5 \times 10^{-11}$eV·m, for exemplifying InSb and InAs-based 2D electron systems [31], unless specified.

We first give the relations among $\Delta_{\mu\nu}$ to make the arguments on the SSE transparent when the system has the $D_2$ symmetry without triangle region. The time-reversal symmetry yields the equality $L^{(1)}_{\mu\sigma,\nu\sigma'} = L^{(1)}_{\nu\bar{\sigma}',\mu\bar{\sigma}}$, which further yields $\Delta_{\mu\nu} = -\Delta_{\nu\mu}$, where $\bar{\sigma}$ means spin reversing [32]. We summarize other relations in Ref. [33]. Figure 2(a) shows $\Delta_{1\nu}$ and $\Delta_{3\nu}$ versus $E_F$ in symmetric systems. We can derive the relations $\Delta_{12} = -\Delta_{32} = \Delta_{34} = -\Delta_{14}$ and $\Delta_{13} = \Delta_{31} = 0$ from the relations given in Ref. [33]. Substituting these into Eq. (5), the SSC yields $S^I_1 = S^I_3 = 0$. These results indicate that there is definitely no longitudinal SSE in the symmetric systems, but the transverse SSE occurs as in the case of Ref. [20] (See Fig. 1). It is straightforward to obtain $S^I_2 = -S^I_4 = \Delta_{23}\hbar/(2e)$.

Figure 2(b) presents the calculated results of $\Delta_{1\nu}$ and $\Delta_{3\nu}$ for asymmetric systems versus $E_F$, which manifests strongly oscillating structures for both of these, originating from energy-dependent transmission coefficients. Peaks appear at step points between two plateaus of the transmission coefficients [20,30]. In comparison with the symmetric system, we find $\Delta_{12}= -\Delta_{32} \neq -\Delta_{14}= \Delta_{34}$ and $\Delta_{13}= -\Delta_{31} \neq 0$. The inequalities ( $\Delta_{12} \neq -\Delta_{14}$ and $-\Delta_{32} \neq \Delta_{34}$ ) and nonzero $\Delta_{13}$ result in non-vanishing $S^I_{\mu=1,3}$ in Eq. (5). Meanwhile, the transverse SSE still exists. Therefore, both the longitudinal SSE and the transverse SSE can be simultaneously realized in our asymmetric system. The realization of the longitudinal SSE originates from the symmetry breaking from the $D_2$ to the $C_2$ with etched triangle region. The physical implications for the above can be made as follows: i) $L^{(1)}_{\mu\sigma,(\mu+1)\bar{\sigma}} = L^{(1)}_{\mu\bar{\sigma},(\mu+1)\sigma}$ given in Ref. [33] does not hold when $\mu$=1,2,3,4; ii) $L^{(1)}_{\mu\sigma,(\mu+1)\bar{\sigma}} = L^{(1)}_{(\mu+1)\bar{\sigma},(\mu+2)\sigma}$ and $L^{(1)}_{\mu\sigma,(\mu+1)\sigma} = L^{(1)}_{(\mu+1)\sigma,(\mu+2)\sigma}$ given in Ref. [33] do not hold when $\mu = 2,4$; iii) $L^{(1)}_{\mu\sigma,(\mu+2)\sigma} = L^{(1)}_{\mu\bar{\sigma},(\mu+2)\bar{\sigma}}$ given in Ref. [33] does not hold when $\mu = 1,3$. Note that $\mu + 2 = 1$ when $\mu = 3$, and $\mu + 1 = 1$, $\mu + 2 = 2$ for $\mu = 4$.

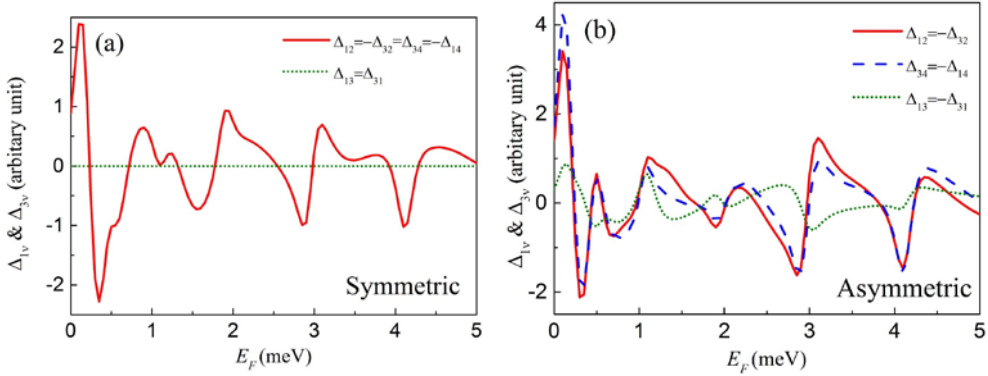

FIG. 2. (Color online) $\Delta_{1\nu}$ ($\nu = 2,3,4$) and $\Delta_{3\nu}$ ($\nu = 1,2,4$) introduced in Eqs. (5) and (6) as a function of Fermi energy in (a) symmetric and (b) asymmetric systems. The inequalities ( $\Delta_{12} \neq -\Delta_{14}$ and $-\Delta_{32} \neq \Delta_{34}$ ) and nonzero $\Delta_{13}$ lead to non-vanishing $S^I_{\mu=1,3}$ in Eq. (5).

Figure 3(a) and 3(b) provide the longitudinal SSCs $\left(S^I_{\mu=1,3}\right)$ and the transverse SSCs $\left(S^I_{\mu=2,4}\right)$ as a function of $E_F$ in both symmetric and asymmetric systems taking $\gamma = 0$. The CB conditions are employed in our calculations. Figure 3(a) for asymmetric system realizes that $S^I_{\mu=1,3}$ oscillate strongly with varying $E_F$. Both of the sign and the amplitude of $S^I_{\mu=1,3}$ oscillate drastically since the transmission coefficients depend highly on $E_F$. The maximum value of the longitudinal SSC becomes $0.0036 k_B$ for $E_F$=0.2meV. In contrast, there exists no longitudinal SSE in symmetric system. Figure 3(b) shows that $S^I_{\mu=2,4}$ oscillate as well with $E_F$ in both symmetric and asymmetric systems. The maximum value of the transverse SSC becomes $0.018 k_B$, five times larger than the longitudinal SSC. The inequality $S^I_2 \neq -S^I_4$ holds for asymmetric system, different from the case $S^I_2 = -S^I_4$ for symmetric system. This indicates that, in symmetric system, the spin current through lead 2 is exactly the same as the spin current through lead 4. There is no spin current in lead 1 and lead 3 which results in the absence of longitudinal SSC. In asymmetric system, the spin current through lead 2 is not equal to the spin current through lead 4. A portion of spin current, aside from the spin currents in the transverse direction, flows into the longitudinal direction yielding nonzero longitudinal SSC as shown in Fig. 1.

Figures 3(c) and 3(d) present the longitudinal SSCs $\left(S^V_{\mu=1,3}\right)$ and the transverse SSCs $\left(S^V_{\mu=2,4}\right)$ in both symmetric and asymmetric systems with the OB conditions. Unlike the CB conditions, $S^V_\mu$ cannot be solved analytically. The numerically calculated $S^V_{\mu=1,3}$ in Fig. 3(c) vary rapidly with $E_F$ in asymmetric system. The extreme value of the longitudinal SSC becomes $-0.038 k_B/e$ for the case of $E_F$=0.45meV. As in the case of the CB conditions, there is no longitudinal SSE in symmetric system, *i. e.* $S_1^V = S_3^V = 0$. Figure 3(d) shows that $S_{\mu=2,4}^V$ oscillate rapidly with $E_F$ in both symmetric and asymmetric systems. The extreme value of the transverse SSC yields $-0.1 k_B/e$, which is larger than the longitudinal SSC. The situation of $S_2^V = -S_4^V$ is realized only in the case of symmetric system, indicating the spin accumulation in lead 2 is exactly the same as that in lead 4 with the reverse sign, and there is no spin accumulation in lead 1 and lead 3. In asymmetric system, the electron spins accumulate in all four leads.

From the results in Figs. 3(a)-(d), we find that spin current or spin accumulation along both the longitudinal and the transverse directions can be generated by applying a temperature gradient when the symmetry along the transverse direction is broken. See Fig. 1 as example, the spin current towards the square region from lead 2 splits into three spin currents out of the square region through leads 1,3,4. This is quite different from the symmetric system in which the spin current or spin accumulation can be generated only along the transverse direction.

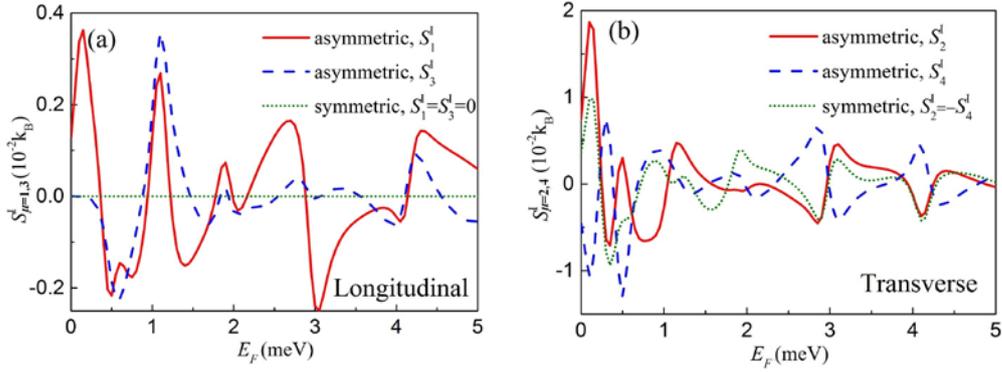

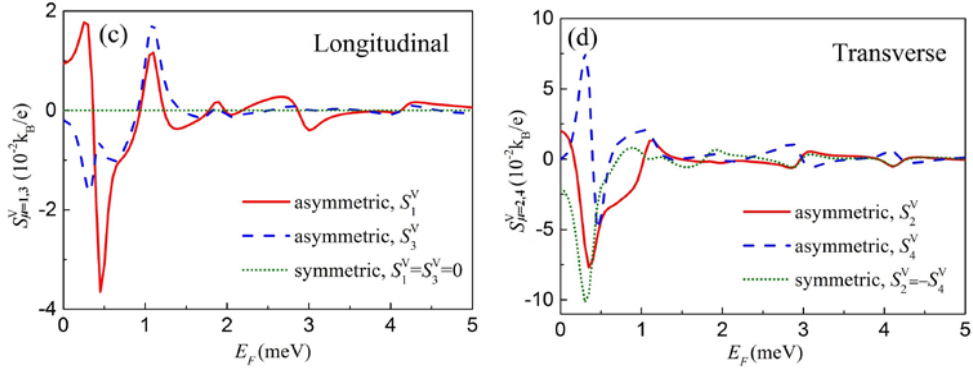

FIG. 3. (Color online) (a) Longitudinal SSC and (b) transverse SSC versus $E_F$ under CB conditions for both symmetric and asymmetric systems. (c) and (d) are the same as (a) and (b) but with OB conditions.

Figure 4(a) shows the calculated results for the longitudinal SSC versus the Rashba SOC constant $V_{SOC}$ under CB conditions in both symmetric and asymmetric systems for $\gamma = 0$ and $E_F = 0.5$meV. In asymmetric system, we find that $S^I_{\mu=1,3}$ vanish for $V_{SOC} = 0$ because of no spin-splitting without the SOC. $S^I_{\mu=1,3}$ vary with increasing $V_{SOC}$ and the amplitudes become larger due to stronger spin-splitting. For some values of $V_{SOC}$ such as $8 \times 10^{-11}$ev·m, not only the amplitude but also the sign of $S^I_1$ is different from $S^I_3$. In symmetric system, there is no longitudinal SSE independent of $V_{SOC}$.

Finally, we consider the effect of the temperature parameter $\gamma$ on the longitudinal SSC. As shown in Fig. 1, $T_1 - T_3 = \Delta T$, $T_1 - T_2 = T_1 - T_4 = \gamma \Delta T$, and $T_2 - T_3 = T_4 - T_3 = (1 - \gamma)\Delta T$. Figure 4(b) shows the longitudinal SSC versus $\gamma$ for the CB conditions in both symmetric and asymmetric systems when $V_{SOC} = 2.5 \times 10^{-11}$eV·m and $E_F = 0.45$meV. In asymmetric system, $S^I_{\mu=1,3}$ vary linearly with $\gamma$. This linear dependence comes from Eq. (5). The absolute value of SSC in lead 1 is larger (smaller) than that in lead 3 for $\gamma < 0.5$ ($\gamma > 0.5$). When $\gamma = 0.5$, $S^I_1$ and $S^I_3$ become the same: $S_1^I = S_3^I = \hbar/(4e)[\Delta_{12} + \Delta_{14} + 2\Delta_{13}]$. It is easy to understand that similar results $S_1^V = S_3^V$, though not presented in this Letter, can be found for the OB conditions when $\gamma = 0.5$. We point out that one can choose $\gamma = 0$ ($\gamma = 1$) to obtain a large SSC in lead 1 (lead 3). Such feature enables us to enhance the longitudinal SSC. In symmetric system, there is no longitudinal SSE whatever how large $\gamma$ is.

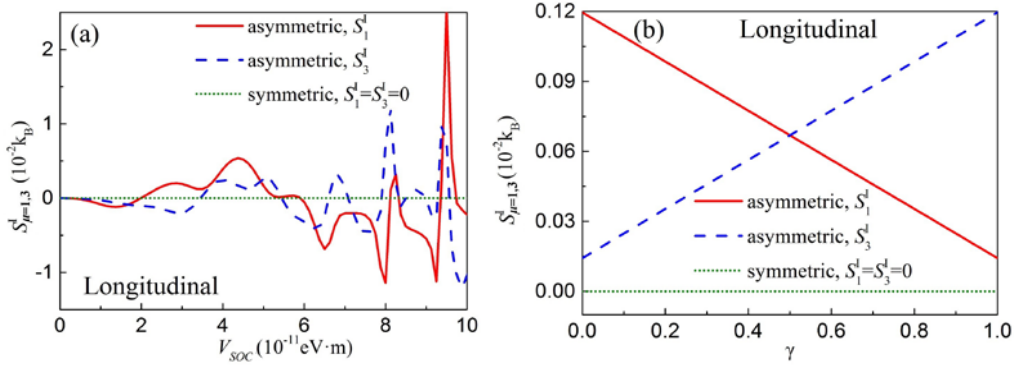

FIG. 4. (Color online) Longitudinal SSC in lead 1 and lead 3 (a) versus the Rashba SOC constant for $\gamma = 0$ and (b) versus $\gamma$ for $V_{SOC} = 2.5 \times 10^{-11}$eV·m. The calculations were performed by taking $E_F = 0.45$meV.

To summarize, we have proposed a new type of the spin Seebeck effect (SSE) emerging from the Rashba spin-orbit coupling in asymmetric four-terminal electron systems, which exhibits both the longitudinal and the transverse SSE in the absence of magnetic fields. Our calculations have revealed that the longitudinal SSE can be realized by the breaking of reflection symmetry along the transverse direction. This is due to that the spin flow along the transverse direction induced by temperature gradient splits into the longitudinal direction by the triangle region. Furthermore, our calculations show that the Fermi energy, the Rashba SOC constant, and the temperature parameter $\gamma$ affect severely the spin Seebeck coefficient. Our findings should be realized in asymmetric four-terminal mesoscopic electron systems, for examples, such as InSb or InAs-based 2D systems with the Rashba SOC.

**Acknowledgements**

JZ, BW, ML, and BL are supported by the National Natural Science Foundation of China Grant No. 11334007. JZ is also supported by the program for New Century Excellent Talents in Universities Grand No. NCET-13-0431. TN acknowledges the support from Grand-in-Aid for Scientific Research from the MEXT in Japan, Grand No. 26400381.